\documentclass[conference]{IEEEtran}
\IEEEoverridecommandlockouts
\usepackage{cite}
\usepackage{amsmath,amssymb,amsfonts}
\usepackage{algorithm,algorithmic}
\usepackage{graphicx}
\usepackage{textcomp}
\usepackage{booktabs}
\usepackage{braket}
\usepackage{tabularx}
\usepackage{xcolor}
\def\BibTeX{{\rm B\kern-.05em{\sc i\kern-.025em b}\kern-.08em
    T\kern-.1667em\lower.7ex\hbox{E}\kern-.125emX}}

\usepackage{titlesec}
\titlespacing*{\section}{0pt}{4pt}{2pt}
\titlespacing{\subsection}{0pt}{4pt}{2pt}

\usepackage{lineno,hyperref}
\hypersetup{
    colorlinks   = false,
    linkcolor    = blue,
    citecolor    = black,
    citebordercolor=green
}
\usepackage{bm}
\usepackage{mathrsfs}
\usepackage{graphicx}
\usepackage{subfigure}
\usepackage[dvipsnames]{xcolor}
\usepackage{nicefrac}
\usepackage{array}
\usepackage{float}
\usepackage{balance}
\usepackage{multirow} 
\usepackage{quantikz}
\usepackage{tikz}
\usetikzlibrary{arrows.meta,shapes,positioning,fit,backgrounds,calc}
\definecolor{borderpurple}{RGB}{128, 55, 155}
\definecolor{lightpurple}{RGB}{233, 225, 242}
\definecolor{medpurple}{RGB}{201, 176, 242}

\usepackage{adjustbox}
\usepackage[flushleft]{threeparttable} 

\makeatletter
\newcommand*{\transpose}{%
  {\mathpalette\@transpose{}}%
}
\newcommand*{\@transpose}[2]{%
  \raisebox{\depth}{$\m@th#1\intercal$}%
}
\newcommand*\diff{\mathop{}\!\mathrm{d}}

\newcommand{\thickhline}{%
    \noalign {\ifnum 0=`}\fi \hrule height 1pt
    \futurelet \reserved@a \@xhline
}
\newcolumntype{"}{@{\hskip\tabcolsep\vrule width 1pt\hskip\tabcolsep}}

\makeatother

\ifCLASSINFOpdf

\else

\fi

\begin{document}
\renewcommand{\ttdefault}{cmtt}
\bstctlcite{IEEEexample:BSTcontrol}


\title{Quantum-Accelerated Deep Reinforcement Learning for Frequency Regulation Enhancement}

\author{\IEEEauthorblockN{Amin Masoumi and Mert Korkali}
\IEEEauthorblockA{\textit{Department of Electrical Engineering and Computer Science}\\ 
\textit{University of Missouri} \\
Columbia, MO 65211 USA \\
e-mail: \{\texttt{am4n5,korkalim\}@missouri.edu}}}

\maketitle

\begin{abstract}
In modern power systems, frequency regulation is a fundamental prerequisite for ensuring system reliability and assessing the robustness of expansion projects. Conventional feedback control schemes, however, exhibit limited accuracy under varying operating conditions because their gains remain static. Consequently, deep reinforcement learning methods are increasingly employed to design adaptive controllers that can be generalized to diverse frequency control tasks. At the same time, recent advances in quantum computing provide avenues for embedding quantum capabilities into such critical applications. In particular, the potential of quantum algorithms can be more effectively explored and harnessed on near-term quantum devices by leveraging insights from active controller design. In this work, we incorporate a quantum circuit together with an ansatz into the operation of a deep deterministic policy gradient agent.
The simulation results of the IEEE 14-bus test system demonstrate the potential of this integrated approach that can achieve reliable, robust performance across diverse real-world challenges.     
\end{abstract}


\section{Introduction}\label{sec:sec1}

Frequency regulation is considered a pivotal practice in response to the frequency incidents in power systems' dynamics, determining how the electrical grid reacts to imbalances between generation and load~\cite{cui2022andes_gym,yusuf2025review}. Hence, reducing the risk of under-frequency load shedding and cascading outages is the result of an effective response that can alleviate the frequency nadir after a large disturbance~\cite{wang2024alleviating,yu2025order}, i.e., load increases and generation trips. To activate the process, the spectral and temporal characteristics of frequency excursions must inform protection settings and tune the automatic generation control (AGC) mechanism, thereby ensuring equitable power sharing~\cite{mohamed2025adaptive}. This mechanism is constructed based on the impact of coefficients of PID controller receiving the area control error (ACE) signals and calculating the gains of powers to be dispatched for the frequency incidents~\cite{chen2025load}. However, these coefficients are static within the AGC operating range, e.g., every 5 seconds. Hence, the probable repercussion can be a delayed response. On the other hand, coordinated multivariable design can pose a challenge for AGC, leading to amplification of oscillations or control conflicts. On the other hand, deep reinforcement learning (DRL) and its variants, as a subcategory of artificial intelligence, demonstrate their capability and flexibility for enhancing frequency response~\cite{zhao2024distributed}. This formidable strategy seeks to bypass the aforementioned obstacle by processing the reflected data to inform the objective functions of the affected environments. By doing so, it starts from a single raw policy and updates the domain of solutions to reach optimality by minimizing the probability of divergence or maximizing the probability of convergence. The authors in~\cite{chen2023investigation} have used deep deterministic policy gradients with a softmax function (SD2) to control the performance of virtual synchronous generators (VSG) under severe conditions. This model-free technique is based on the DDPG's functionality, leveraging online actor and critic capabilities. In this regard, the SD2 agent attempts to restore post-fault rotor stability by adjusting the virtual inertia and the reference active power. Hence, the reward function penalizes instability and angular-frequency deviation in the system. However, the speed of parameter optimization, i.e., the gradient calculation for the weight and bias, is constrained by policy exploration, which simultaneously affects the accuracy of the actor and critic. As the environment becomes more complex, convergence can take longer. In \cite{eskandari2023deep,oboreh2023virtual,lee2024deep}, deep deterministic policy gradient (DDPG), twin delayed deep deterministic policy gradient (TD3), and softmax DDPG have been developed to control a GFM converter. The main role of the conducted DDPG is to emulate virtual inertia through impedance sharing under time-varying load conditions. The target of the reward function is to increase the frequency response of the IBR-rich grid with respect to the rate of change of frequency (RoCoF), angular frequency, etc. Reference~\cite{ma2022blockchain} proposes a DDPG algorithm to serve as the
demand response mechanism in the smart grid. Next, the DDPG algorithm
is trained in a centralized manner, utilizing a global state
that encompasses information on all the control strategies
involved in executing demand response. Furthermore, the energy management issue has
been formulated as an MDP in~\cite{ye2020model} using prioritized deep deterministic
policy gradient (PDDPG), which makes use of DRL’s
capabilities to establish real-time autonomous control methods
for the multi-energy system. The goal of the DRL approach
is to tackle the problems associated with multi-dimensional
continuous state and action spaces in energy management. By
investigating the action space, deep neural network (DNN) can find more economical
solutions and obtain precise feedback about how its energy
management decisions affect the target environment. However, the high-dimensionality~\cite{masoumi2025adversarially} of state and action spaces can
cause scalability issues, making it difficult for the DRL
agent to explore an effective range of policies and converge with the granularity of the solutions. In this regard, quantum mechanics possesses a trait that
can address these obstacles by facilitating the exploration in the solution stage tools that
can suppress complexity~\cite{wu2025quantum}. The main philosophy of using a
quantum perspective is to leverage its properties (i.e., superposition
and entanglement) to boost the performance of DRL agent~\cite{weidner2025robust}. The subtle notion here is that the application of ansatz or
a parameterized quantum circuit (PQC)~\cite{chen2026quantum}that can learn the interaction of computing units, e.g., actor and critic, and enact to increase the convergence speed of the agent~\cite{wu2025quantum,kruse2025benchmarking}. Motivated by these considerations, this paper proposes a methodology for developing a quantum-enhanced DRL algorithm tailored to frequency response, with the aim of improving operational planning in power systems. Furthermore, it systematically investigates and delineates the broad landscape of quantum-accelerated DRL use cases within the complex and evolving context of the electric utility industry.
Each component adds to the overall story. Our contributions are highlighted as follows:

\begin{itemize}
    \item Introducing an exact formulation for accelerating the performance of actor and critic networks,
    \item Developing a detailed framework for addressing the high-dimensionality of DRL via PQCs; and
    \item Designing accurate, fast frequency responses based on the impact of the incident 
\end{itemize}

The remainder of this paper is organized as follows: Section \ref{sec:sec2} constructs the conventional frequency regulation formulation and discusses the functionality of the AGC mechanism based on area control error signals. Section \ref{sec:sec3} introduces the proposed framework for integrating PQC with DDPG by constructing ansatz for actor and critic updates. Sections \ref{sec:sec4} and \ref{sec:sec5} present the case study and the results of the validation of the proposed method in the IEEE 14-bus test system. Finally, Section \ref{sec:sec6} concludes the paper and provides future directions.

\section{Frequency Regulation via Automatic Generation Control}\label{sec:sec2}
The overarching objective of AGC is to maintain the equilibrium of interconnected power networks by monitoring power mismatches across tie lines and disturbances in system frequency. In addition, system stability is regulated by continuously adjusting the dispatch of generating units. In this way, AGC supplements the governor's response, mitigating deviations from the nominal system frequency. The residual frequency error is eliminated by using ACE signals, which maintain a long-term equilibrium between power generation and electrical load.


To implement this regulation in a coordinated manner, AGC delegates control signals to a selected set of regulating units using the PI controller scheme via
\begin{equation}
\label{eq:eq1}
\text{ACE}(t) = \Delta P_{\text{tie}}(t) - 10B \cdot \Delta f (t),
\end{equation}

\noindent where $\Delta P_{\text{tie}}$, and $B$, $\Delta f$,  are the power interchange of the tie lines, frequency bias, and frequency deviation from the actual frequency, respectively~\cite{R41}. The AGC signal is calculated based on the PI controller and ACE signals as
\begin{subequations}
\begin{align} 
 U(s) &= -\Big(K_P + \frac{K_I}{s}\Big)\,\text{ACE}(s),  \label{eq:2}\\
 u(t) &= -K_P\,\text{ACE}(t) - K_I \int_{0}^{t} \text{ACE}(\tau)\,\diff\tau \label{eq:eq3}, 
\end{align}
\end{subequations}



\noindent where $u(t)$ is the total AGC command [MW] to be distributed among participating generators, and $K_P, K_I\ge 0$ are controller gains. The sending signals aim to activate the ramping capability and economic dispatch priorities, i.e., the share of each generator. Hence, the commanded signal changes the power set point of each Generator $j$ as follows:
\begin{subequations}
\begin{align} 
 \Delta P_{g_j}^{\text{cmd}}(t) &= \alpha_j\,u(t)  \label{eq:4}\\
 \sum_{j=1}^{N} \alpha_j &= 1,\qquad 0\le \alpha_j \le 1 \label{eq:eq5}, 
\end{align}
\end{subequations}



\noindent where $\Delta P_{g_j}^{\text{cmd}}(t)$ and $\alpha_j $ are the rescheduled power set point and the share of each generator, respectively.

\section{Proposed Quantum-Accelerated DRL}\label{sec:sec3}
In this section, the architecture of quantum-accelerated DRL is described in detail.

\subsection{Fundamentals of DDPG}
In this section, we review the functionality of the DPPG algorithm, a variant of actor-critic networks trained in an off-policy manner. This technique aims to provide continuous-action control by combining the traits of deterministic policy gradients with value-based learning, specifically a Q-network. Consequently, the actor network ($\mu(s|\theta^\mu)$) is responsible for devising long-term rewards by selecting the best deterministic action in its maximization problem. On the other hand, the critic network ($Q(s,a|\theta^Q)$) has the role of reviewer in assessing the value of each state–action pair produced by the actor, in which $\theta^\mu$ and $\theta^Q)$ are the parameters of the actor and critic updated during training. Hence, the actor tries to map states or environment observations to the continuous action space and is audited by the critic in each state, which helps to learn the value of the state-action pair. Moreover, DDPG maintains separate target networks, represented by $\mu'(s|\theta^{\mu'})$ and $Q'(s,a|\theta^{Q'})$, for both the actor and the critic, which improves the stability of the training. These targets track the primary networks through a slow slope of parameter updates and block oscillations. Generally, the deterministic policy gradient theorem is deployed to update the parameters of the actor network by maximizing the expected return:
\begin{equation}
   \label{eq:eq6} 
   \nabla_{\theta^\mu} \mathbb{J} \approx \mathbb{E}_{s_t \sim \rho^\beta} \left[ \nabla_a Q(s,a|\theta^Q)|_{s=s_t,a=\mu(s_t)} \nabla_{\theta^\mu} \mu(s|\theta^\mu)|_{s=s_t} \right].
\end{equation}

\noindent where $\rho^\beta$ expresses the distribution of the state according to the behavior of the policy taken $\beta$. Hence, the critic's role is to direct the actor to obtain actions that yield a higher estimated value with
\begin{subequations}
\begin{align} 
 \mathbb{L}(\theta^Q) &= \mathbb{E}_{s_t,a_t,r_t,s_{t+1} \sim \mathcal{D}} \left[ \left( Q(s_t,a_t|\theta^Q) - y_t \right)^2 \right] \label{eq:eq21}\\
 y_t &= r_t + \gamma Q'(s_{t+1}, \mu'(s_{t+1}|\theta^{\mu'})|\theta^{Q'}). \label{eq:eq22} 
\end{align}
\end{subequations}

The key subtlety lies in the fact that the sampled experience tuple, $(s_t, a_t, r_t, s_{t+1})$, is drawn from the replay buffer $D$, while the bootstrapping step relies on $y_t$ to compute the target value, which is, in turn, determined by slowly updated target networks.
Note that the target networks follow soft-update rules with parameter $\tau \ll 1$ (typically around $0.005$) ensures gradual tracking as
\begin{align}
\label{eq:eq23}
\theta^{Q'} &\leftarrow \tau \theta^Q + (1-\tau)\theta^{Q'} \\
\theta^{\mu'} &\leftarrow \tau \theta^\mu + (1-\tau)\theta^{\mu'}
\end{align}

\subsection{Quantum Computing Preliminaries}
Quantum computing is a novel paradigm that leverages the principles of quantum mechanics to process information in Hilbert space. The conceptual idea is based on two orthonormal basis states. 
\begin{equation}\label{eq:eq6}    
\ket{0} = \begin{bmatrix} 1 \\ 0 \end{bmatrix}, \quad \ket{1} = \begin{bmatrix} 0 \\ 1 \end{bmatrix}.  
\end{equation}

The computational power comes from the quantum principles of superposition, entanglement, and interference. Therefore, an arbitrary state, $\ket{\psi}$, of a solution is a linear combination (superposition) of these basis states, that is,
\begin{equation}\label{eq:eq7} 
\ket{\psi} = \alpha \ket{0} + \beta \ket{1}, |\alpha|^2 + |\beta|^2 = 1.
\end{equation}

The entanglement can be expressed as a tensor product of individual qubits as
\begin{equation}\label{eq:eq7} 
\ket{\Phi} = (\alpha\ket{0} + \beta\ket{1}) \otimes (\gamma\ket{0} + \delta\ket{1}).    
\end{equation}

Inference is the process of extracting probability $p(m) = \bra{\Phi} M^\dagger M \ket{\Phi}$, information from Measurements $M$ of the quantum circuit orchestrated by the superposition and entanglement stages, given a specific state $\Phi$ as
\begin{equation}\label{eq:eq8}
\frac{M \ket{\Phi}}{\sqrt{\bra{\Phi} M^\dagger M\ket{\Phi}}},  
\end{equation}

\noindent where $M$ is the Pauli unitary gates that can be selected based on the Hamiltonian of the system.

\subsection{The Architecture of Quantum-Accelerated DDPG}
The DDPG framework relies fundamentally on two key quantum-mechanical principles introduced earlier: superposition and entanglement. These concepts constitute the primary motivation for embedding PQCs in the update routines of both the actor and critic networks. Within this quantum setting, two central questions arise: (1) \textit{how can one exploit the expressive capacity of quantum state representations?} and (2) \textit{how can one encode inter-state correlations to drive parameter updates?}
To initiate the procedure, the superposition provides expressiveness by allowing qubits to occupy multiple basis states simultaneously. Therefore, the solution space of the underlying control problem can be better explored by testing a variety of circuit connections. Moreover, the entanglement stage complements the task by representing the dynamics of a quantum system through correlations between superpositions of distinct quantum states, as shown in Algorithm~\ref{alg:PQC_DDPG}. More specifically, the encoding prepares the classical states of observation by applying Pauli gates to convey meaningful functionality in the quantum realm. According to this logic, the proposed algorithm aims to make great use of the $R_Y$, $R_Z$, and $CNOT$ operators. To tune single-qubit rotations, $R_Y$ and $R_Z$ operators are utilized to facilitate the conversion of classical states to quantum states, whereas the $CNOT$ gate introduces entangling interactions. As shown, these operators form the foundation of the quantum circuit required for the update steps of both the actor and critic networks. Consequently, a single unitary layer in the PQC is an architecture as a sequence of input-dependent unitary rotations that can be updated based on the trainable values (angles) as follows:
\begin{equation}\label{eq:eq9}
\begin{aligned}
U_\ell(x;\alpha,\beta,\theta)
&= E \circ
\Bigg(\bigotimes_{j=0}^{n-1} R_Y(\theta_{\ell,j}^{(2)})\,R_Z(\theta_{\ell,j}^{(1)})\Bigg) \\
&\quad \circ\;
\Bigg(\bigotimes_{j=0}^{m-1} R_Z(\beta_{\ell,j} x_j)\, R_Y(\alpha_{\ell,j} x_j)\Bigg),
\end{aligned}
\end{equation}

\noindent where $x$, $\alpha$, $\beta$, $\theta$, and $E$ denote the classical inputs and the corresponding scaling parameters, the trainable rotation parameters, and the entangling block ($CNOT$), respectively. The overall circuit generates a quantum state by stacking $L$ layers of repetition as
\begin{equation}
    \label{eq:eq10}
\ket{\psi} = U_L\,U_{L-1}\cdots U_1\,\ket{\mathbf{0}},
\end{equation}
and the circuit’s complete unitary map may be written compactly as
\begin{equation}
\label{eq:eq11}
U(\theta) = \prod_{l=1}^{ L} U_{\text{ent}} U_{\text{rot}}(\theta_l) U_{\text{encode}}(x),
\end{equation}

\noindent where $U_{\text{encode}}(x)$, $U_{\text{rot}}(\theta_l)$, and $U_{\text{ent}}$ are deployed to provide the encoding purpose of classical states into quantum states, constructing trainable rotations, and correlating entanglement between different qubits throughout Layers $L$. 
\subsection{Quantum Integration of Actor Network}

Now that we have described PQC as the tensor product of Pauli unitary gates, we must employ them in the actor network's update steps. In this regard, the expectation values of the Pauli-Z measurements serve as the integrated PQC to construct a quantum policy $\mu_Q(s|\theta^\mu)$.
\begin{equation}
\label{eq:eq12}
\mu_Q(s) = f_{\text{post}}(  \bra{0} U^\dagger(s, \theta^\mu) Z_i U(s, \theta^\mu) \ket{0} ),
\end{equation}
where $U(s, \theta^\mu)$ and $f_{\text{post}}$ are the deployed PQC ansatz and the classical post-processing function to determine the action in response to the varying dynamics of the power system, respectively, and $U^\dagger$ denotes the complex conjugate of $U$. In addition, $Z_i$ are the Pauli-Z measurement operators that are instructed to be deployed in all qubits. In this regard, the quantum actor implements the mapping via $f_{\text{post}}$ as
\begin{equation}
\label{eq:eq13}
\mu_Q(s) = A \cdot \tanh\left( \frac{1}{K} \sum_{i=1}^{N_a} \braket{Z_i} \right) + B,
\end{equation}
where $A$ and $B$ are the scaling factors of action states and bias vectors, respectively. Moreover, $N_a$, $K$, and $\braket{Z_i}$ are the number of actions, normalization constant, and expectation values from quantum measurements, respectively. Now that we have devised the PQC step of the actor network, the gradient of the updates regarding the actor networks is given by
\begin{subequations}\label{eq:eq14}
\begin{align}
\frac{\partial \mu_Q(s)}{\partial \theta_i}
&= \tfrac{1}{2}\!\left[\mu_Q\!\left(s;\,\theta_i + \tfrac{\pi}{2}\right)
      - \mu_Q\!\left(s;\,\theta_i - \tfrac{\pi}{2}\right)\right]
\label{eq:25a} \\[4pt]
\nabla_{\theta^\mu} \mathbb{J}
&\approx \mathbb{E}_{s \sim \mathcal{D}}
\!\left[\, \nabla_a Q_Q\!\left(s,a \mid \theta^Q\right)\big|_{a=\mu_Q(s)}\;
          \nabla_{\theta^\mu}\mu_Q\!\left(s \mid \theta^\mu\right) \right].
\end{align}
\end{subequations}

\begin{algorithm}
\caption{Integration of Parameterized Quantum Circuits (PQCs) into the DDPG Framework}
\label{alg:PQC_DDPG}
\begin{algorithmic}[1]
\renewcommand{\algorithmicrequire}{\textbf{Input:}}
\renewcommand{\algorithmicensure}{\textbf{Output:}}
\REQUIRE Classical State $s$, Action $a$; PQC parameters $\theta_\mu, \theta_Q$; encoding scales $\alpha, \beta$; entangling map $E$; replay buffer $D$
\ENSURE Quantum Actor $\mu_Q(s)$ and Quantum Critic $Q_Q(s,a)$

\textit{Initialization}: 
\STATE Initialize PQC layers $U_\ell$, $\ell = 1,\dots,L$
\STATE Initialize target parameters $\theta_\mu', \theta_Q'$
\STATE Set learning rates $\eta_\mu, \eta_Q$

\FOR{each training iteration}
    \STATE Sample minibatch $(s,a,r,s')$ from replay buffer $D$

    \STATE \textbf{Quantum Encoding:}
    \FOR{each input component $x_j \in s$}
        \STATE Apply rotations $R_Y(\alpha_j x_j)$, $R_Z(\beta_j x_j)$
    \ENDFOR

    \STATE \textbf{Construct PQC State:}
    \FOR{$\ell = 1:L$}
        \STATE Apply trainable rotations $R_Z(\theta_{\ell,j}^{(1)})$, $R_Y(\theta_{\ell,j}^{(2)})$
        \STATE Apply entangling layer $E$
    \ENDFOR
    \STATE Obtain final quantum state $|\psi\rangle = U_L \dots U_1 |0\rangle$

    \STATE \textbf{Quantum Actor Output:}
    \STATE Measure Pauli-$Z$ expectation values $\langle Z_i \rangle$
    \STATE Compute action:
    \[
        \mu_Q(s) = A \cdot \tanh\!\left(\frac{1}{K}\sum_{i=1}^{N_a} \langle Z_i \rangle\right) + B
    \]

    \STATE \textbf{Actor Gradient :}
    \FOR{each parameter $\theta_i \in \theta_\mu$}
        \STATE Compute:
        \[
        \frac{\partial \mu_Q(s)}{\partial \theta_i}
        = \frac{1}{2}\left[
            \mu_Q(s;\theta_i+\frac{\pi}{2}) -
            \mu_Q(s;\theta_i-\frac{\pi}{2})
        \right]
        \]
    \ENDFOR

    \STATE \textbf{Quantum Critic Output:}
    \STATE Encode $[s,a]$ into PQC and build $|\psi_{Q}\rangle$
    \STATE Measure Pauli-$X$:
    \[
        Q_Q(s,a) = w_{\text{out}}\langle X_0 \rangle + b_{\text{out}}
    \]

    \STATE \textbf{Critic Loss:}
    \[
        L(\theta_Q) =
        (Q_Q(s,a) - [r + \gamma Q_Q'(s', \mu_Q'(s'))])^2
    \]

    \STATE Update critic: $\theta_Q \leftarrow \theta_Q - \eta_Q \nabla_{\theta_Q} L$
    \STATE Update actor: $\theta_\mu \leftarrow \theta_\mu + \eta_\mu \nabla_{\theta_\mu} J$

    \STATE $\theta_\mu' \leftarrow \tau \theta_\mu + (1-\tau)\theta_\mu'$
    \STATE $\theta_Q' \leftarrow \tau \theta_Q + (1-\tau)\theta_Q'$

\ENDFOR

\RETURN Quantum policy $\mu_Q$ and quantum critic $Q_Q$

\end{algorithmic}
\end{algorithm}

\subsection{Quantum Integration of Critic Network}

For the critic, the PQC is similarly applied, but with a Pauli-X observable to evaluate the value of state–action pairs:
\begin{equation}
\label{eq:eq15}
Q_Q(s,a) = w_{\text{out}} \cdot \bra{0} U^\dagger([s,a], \theta^Q) X_0 U([s,a], \theta^Q) \ket{0} + b_{\text{out}},
\end{equation}

\noindent where $w_{\text{out}}$ and $b_{\text{out}}$ express the trainable output scaling parameters. In addition, $[s,a]$ is used to represent the concatenation of state and action. Here, the observable is the Pauli-X measurement on the first qubit, which captures the Q-value of the state-action pair. In other words, the critic ansatz stacks the single observation of the expected value of the first qubit into a tensor. This value is multiplied by the output scaling parameters. In doing so, the critic update is rewritten as
\begin{equation}\label{eq:eq16}
\begin{aligned}
\mathbb{L}(\theta^Q)
&= \mathbb{E}_{(s,a,r,s') \sim \mathcal{D}} \Bigg[ \Big(
    Q_Q(s,a\mid\theta^Q) \\
&\qquad\qquad - \big(r + \gamma\, Q'_Q\!\big(s',\, \mu'_Q(s'\mid\theta^{\mu'}) \mid \theta^{Q'}\big)\big)
\Big)^2 \Bigg].
\end{aligned}
\end{equation}

\section{Case Study Description }\label{sec:sec4}

In this paper, the load-frequency problem is studied using quantum-accelerated DDPG on the IEEE 14-bus test system. The frequency incident is the $60\%$ load increase that occurs at $t = 1$ s and on Bus 4. The SG frequency is considered the observation, and the power setpoints of SGs after economic dispatch are the actions. The procedure is implemented using the \texttt{Andes\_gym} library~\cite{cui2022andes_gym}. For the configuration of DDPG, Python 3 and the CleanQRL library~\cite{kruse2025cleanqrl} were used. The quantum framework of the PQC is implemented through the PennyLane library~\cite{bergholm2018pennylane}. Furthermore, the actor's PQC comprises 5 qubits, achieved by substituting 5 SGs, and spans five observation spaces, whereas the critic incorporates 10 qubits for state-action pair assessment.

\section{Results and Discussions}\label{sec:sec5}

In this section, the simulation results are presented based on the application of the implemented quantum ansatz, targeting efficiency and adequacy in response to a frequency incident. It must be noted that the training of our DDPG agent begins at Episode 20 and the 400th time step to model the delay in the exploration noise. The main functionality of the pursued algorithm is the application of ansatz that defines the expressibility and trainability of quantum models. Specifically, the hardware-efficient ansatz (HEA) must address the practical constraints of noisy intermediate-scale quantum (NISQ) devices through native gate-set compatibility. In other words, native unitary operators constitute a hard constraint for near-term devices during the agent's training. The interplay between the HEA design and gradient computation is an adequate criterion. Hence, the best parameters (e.g., angles) of the HEA are determined using a differentiation method that probes the solution space. Here, we have applied two perspectives: ``adjoint'' and the ``parameter-shift rule.'' The relationship lies in how HEA's design affects the gradient landscape. Consequently, the agent's training procedure follows the same logic and demonstrates an advantage over the course of policy learning.

\begin{figure}[htbp]
\centerline{\includegraphics[width=\linewidth]{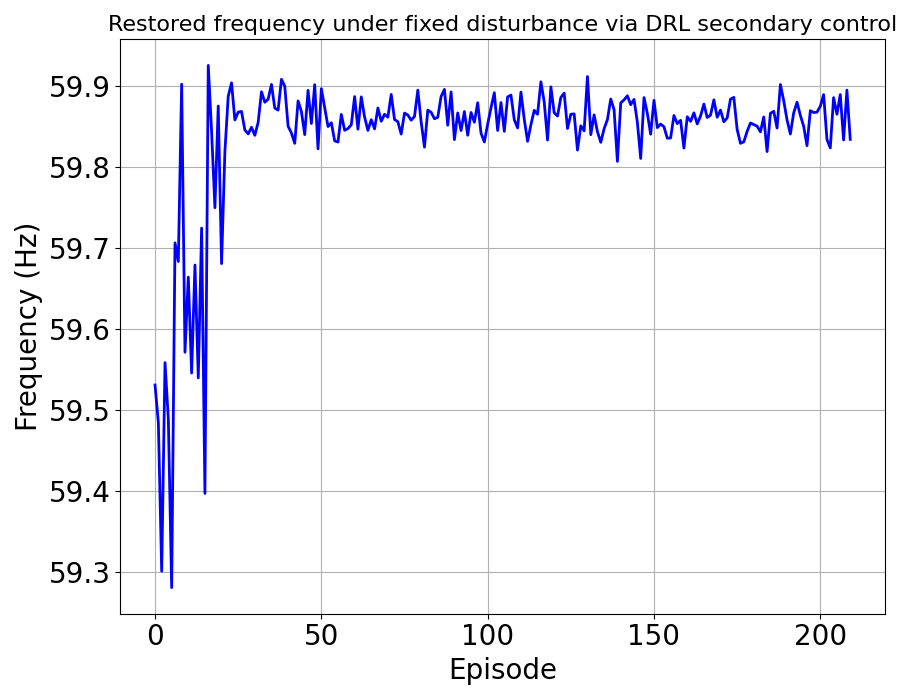}}
    \vspace{-.1cm}
    \caption{The training process of the agent via the adjoint method.}
    \label{fig:fig1}
\end{figure}

\begin{figure}[htbp]
\centerline{\includegraphics[width=\linewidth]{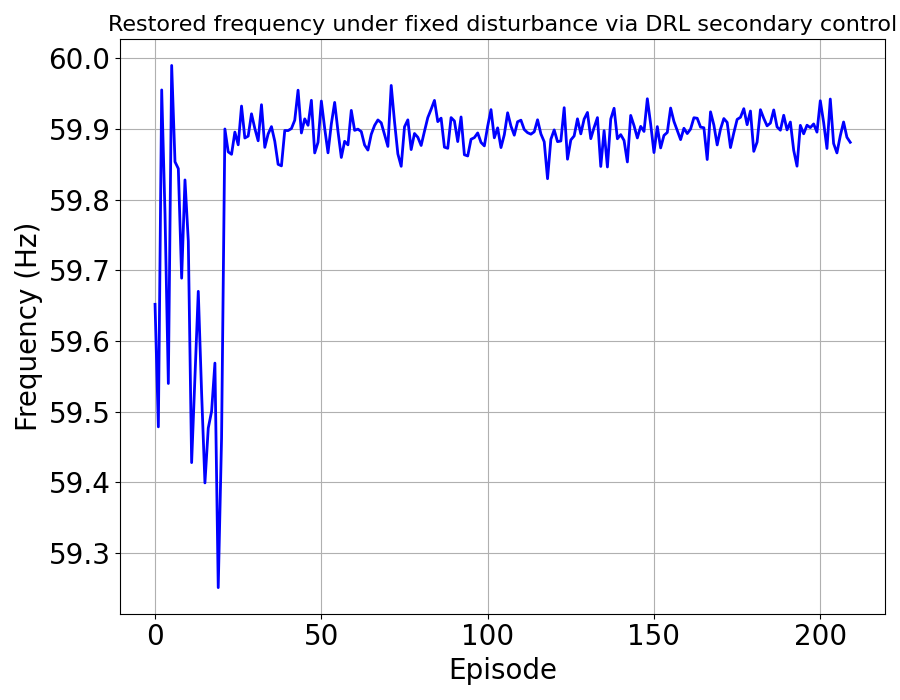}}
    \vspace{-.1cm}
    \caption{The training process of the agent via the parameter-shift rule.}
    \label{fig:fig2}
\end{figure}

\begin{figure}[htbp]
\centerline{\includegraphics[width=\linewidth]{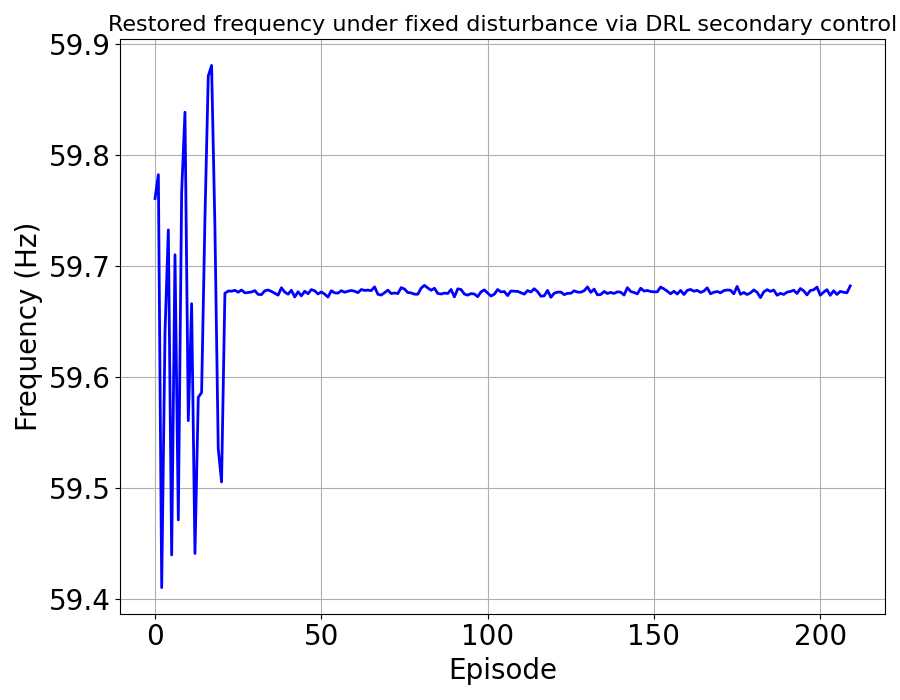}}
    \vspace{-.1cm}
    \caption{The training process without exploration noise}
    \label{fig:fig3}
\end{figure}

According to Figs.~\ref{fig:fig1} and \ref{fig:fig2}, the parameter-shift rule outperforms the adjoint method and demonstrates a better frequency response by maintaining the frequency above $59.9$ Hz. Hence, parameter-shift rule-oriented design can address the computational overhead inherent in quantum gradient methods. Note that the choice of gradient also has a trace of compatibility with the tensor product of the superposition and entanglement steps. The reason is the impact of noise in the NISQ era, which can threaten the stability of training. In other words, coherence times and gate fidelities are the hard constraints of training on NISQ devices. In this regard, canceling noise in the agent's learning procedure cannot help much in alleviating this issue. As shown in Fig.~\ref{fig:fig3}, neglecting exploration noise results in inadequate training, hindering the agent's performance.

On the other hand, executing multiple circuits always violates the balance between circuit depth and complexity (rotation gate choice). Furthermore, we have applied IBM quantum backends to examine the impact of such noisy conditions in practice. In this regard, we have trained the agent using a fake-backend simulator that estimates the performance of an NISQ device, including its noise characteristics. It mimics the noise effects and error-mitigation steps of a real quantum computer, which are critical for the real-world application of training a DRL agent.

\begin{figure}[htbp]
\centerline{\includegraphics[width=\linewidth]{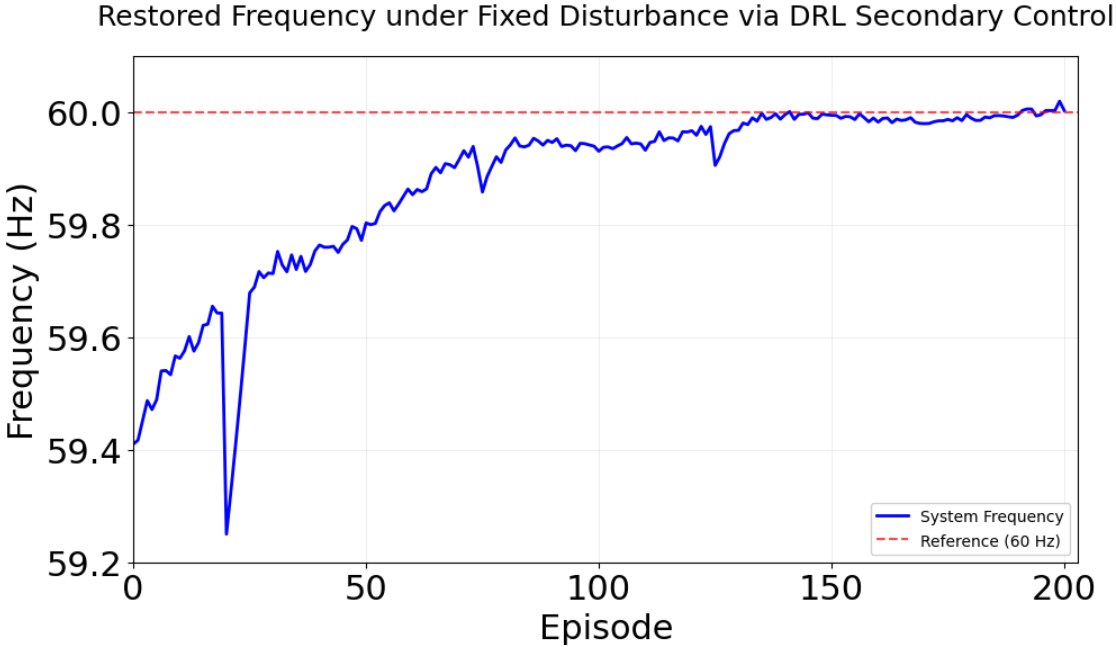}}
    \vspace{-.1cm}
    \caption{The training process quantum-accelerated agent }
    \label{fig:fig4}
\end{figure}

According to Fig.~\ref{fig:fig4}, the agent successfully captures the frequency incident by responding throughout the training episodes and reaching the ideal $60$ Hz. The subtle point here is the agent's continuous improvement during training episodes. This is a valid indicator of the agent's performance in NISQ era. We have also conducted a sensitivity analysis to test the adequacy of the implemented ansatz in response to the variations in the circuit hyperparameters.

\begin{figure}[htbp]
\centerline{\includegraphics[width=\linewidth]{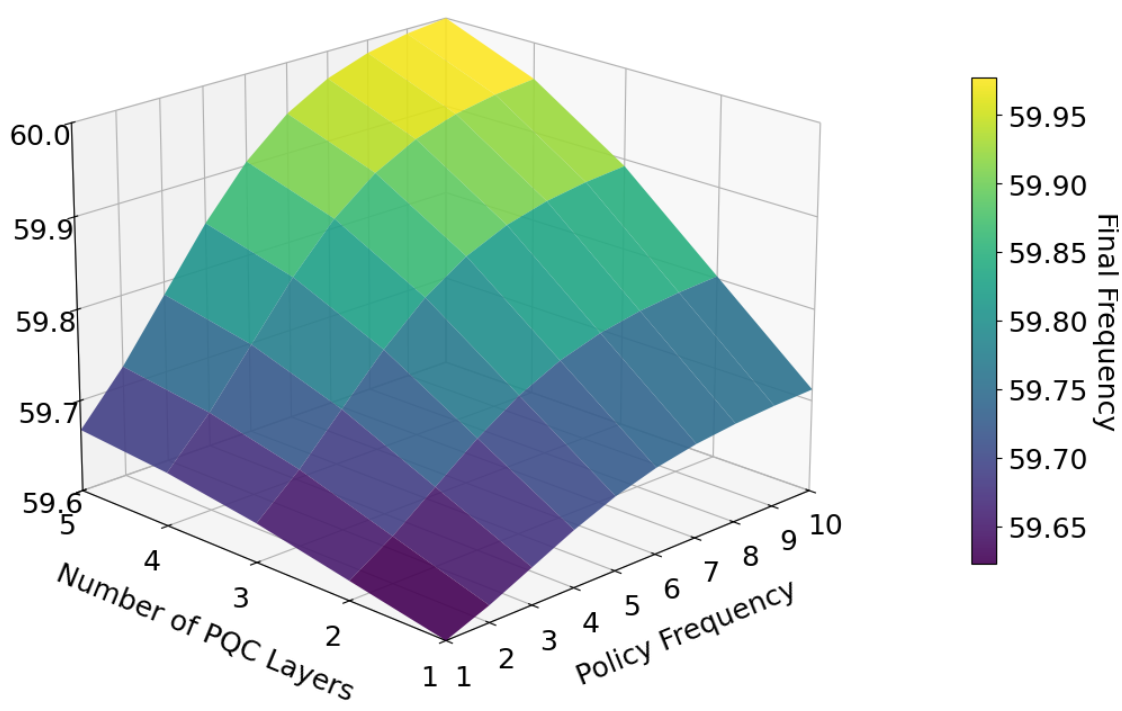}}
    \vspace{-.1cm}
    \caption{The training process of the agent via the parameter-shift rule.}
    \label{fig:fig5}
\end{figure}

According to Fig.~\ref{fig:fig5}, the value of the final frequency increases with the number of PQC repetitions and frequency of policy changes. It is worth noting that the final frequency is more sensitive to the policy update during training. It is because the PQC circuit serves the expressibility purpose in the process, but it has limitations.  Hence, Fig.~\ref{fig:fig5} demonstrates the desired balance between circuit depth and complexity that can satisfy the adequacy constraint in the HEA. This demonstrates the simulation's reliability by accelerating the agent's performance in the NISQ era.


\section{Concluding Remarks}\label{sec:sec6}

In this paper, the quantum-accelerated DDPG is applied to a fixed-frequency incident. The methodology employed captures the stated objective by integrating PQC into the training of actor and critic networks. The case study is applied to the IEEE 14-bus test system. The results demonstrate the efficiency and adequacy of the studied ansatz with respect to the hard constraints of HEA in NISQ devices. In this regard, it has been shown that hardware-aware design exhibits acceptable behavior by minimizing unnecessary parameterization in the PQC. Hence, the outcome of this synergy paves the way for stable training and better convergence. The pursued quantum agent captures nuanced control policies by bridging the gap between theoretical expressibility and real-world application on near-term quantum hardware.


\ifCLASSOPTIONcaptionsoff
\newpage
\fi

\bibliographystyle{IEEEtran}
\bibliography{IEEEabrv,References}

\end{document}